\documentclass[prd,nofootinbib,showpacs,showkeys]{revtex4}
\usepackage{amsmath,graphicx}

\begin{document}

\title{There is no local chiral perturbation theory at finite temperature}

\author{Stefan Leupold}


\affiliation{Institut f\"ur Theoretische Physik, Justus-Liebig-Universit\"at
Giessen, Germany}

\begin{abstract}
For a low-temperature expansion in QCD it is well-known that the Lagrangian of
{\it vacuum} chiral perturbation 
theory can be applied. This is due to the fact that the thermal effects of the heavy
modes are Boltzmann suppressed. The present work is concerned with 
the situation of higher
temperatures, but still below the chiral transition. For a systematic approach it
would be desirable to have a temperature modified chiral perturbation theory at hand
which would yield a chiral power counting scheme. It is shown that this cannot
be obtained.
In principle, chiral perturbation theory emerges from QCD by (a) integrating 
out all degrees of freedom besides the Goldstone bosons and (b) expanding the
obtained non-local action in terms of derivatives. It is shown that at finite
temperature step (b) cannot be carried out due to non-analyticities appearing for
vanishing four-momenta. Therefore one cannot obtain a 
local Lagrangian for the Goldstone boson fields with
temperature modified coupling constants. 
The importance of the non-analyticities is estimated.
\end{abstract}


\maketitle

\section{Introduction}
\label{sec:intro}

Chiral perturbation theory ($\chi$PT) 
has become an important tool to describe in a systematic
way low energy QCD 
(see e.g.~\cite{gasleut1,gasleut2,Ecker:1995gg,Pich:1995bw,Scherer:2002tk}). 
The spontaneous breakdown of chiral symmetry causes the appearance of 
(quasi-)Goldstone bosons which are much lighter than all other hadron 
species.\footnote{Due to the small explicit breaking of chiral symmetry by the finite
quark masses we do not have Goldstone bosons, i.e.~massless modes, in the strict sense.}
In addition, chiral symmetry breaking enforces derivative couplings of the Goldstone
bosons to themselves (and to all other hadrons) --- besides terms which scale with
the small mass of the Goldstone bosons.
The gap in the excitations and the weakness of the couplings at small energies (and
momenta) causes the success of $\chi$PT as a systematic approach
to QCD (at low energies). To describe the interaction of the Goldstone bosons with
themselves and with external sources (e.g.~with photons) the effective Lagrangian of 
$\chi$PT is given by
\begin{equation}
  \label{eq:chiptweinb}
{\cal L} = {1 \over 4} \, F^2 \, {\rm tr}\left(
D_\mu U  D^\mu U^\dagger + \chi U^\dagger + \chi^\dagger U 
\right) \quad \mbox{+ higher order terms.}
\end{equation}
Corrections come with higher orders in powers of the involved energies and the Goldstone
boson masses. Hence for sufficiently small energies these corrections are small and
can be systematically taken into account up to the desired order. 
In (\ref{eq:chiptweinb}) the Goldstone boson fields are encoded in the flavor matrix
$U$. The masses of the Goldstone bosons as well as external scalar and pseudo-scalar
fields are contained in $\chi$, derivatives as well as external vector and axial-vector 
fields in $D_\mu$ 
(cf.~e.g.~\cite{gasleut1,gasleut2,Ecker:1995gg,Pich:1995bw,Scherer:2002tk} for details).

In principle, the coupling constants of $\chi$PT like $F$ given in (\ref{eq:chiptweinb})
can be obtained from QCD by integrating out all degrees of freedom besides the Goldstone
bosons. Such a task, however, would be more or less equivalent to solving QCD in the
low-energy regime. Lattice QCD \cite{Creutz:1984mg} has started to determine some of 
these low-energy constants (see e.g.~\cite{Giusti:2004yp} and references therein). 
In practice, one determines these coupling
constants from experiment \cite{gasleut1,gasleut2,Amoros:2001cp}
or from hadronic \cite{eckgas,Ecker:1989yg,Donoghue:1989ed,pelaez02} 
or quark models \cite{diakpet,espraf,Schuren:1992sc,Mueller:1994dh,Pich:1995bw}.

All the previous statements describe vacuum physics. In principle, it would be nice if 
the same worked at finite temperature $T$. For the following considerations we shall
split the temperature region in three parts:
\begin{itemize}
\item[I.] Low temperatures: All thermal effects caused by the non-Goldstone modes
are suppressed by Boltzmann factors $e^{-M_h/T}$ where $M_h$ is the mass of a
non-Goldstone mode. At low temperatures one can neglect these effects. In this regime
the Lagrangian (\ref{eq:chiptweinb}) of vacuum $\chi$PT can be applied 
\cite{Goity:1989gs,Gerber:1989tt,Gerber:1990yb,Schenk:1993ru}. 
Of course, thermal propagators for the Goldstone bosons are used,
but the input Lagrangian remains unaltered.
\item[II.] Intermediate temperatures: 
In this regime the non-Goldstone modes become important.
It ranges up to the chiral transition.
\item[III.] High temperatures: Beyond the chiral and deconfinement transition quarks
and gluons become the relevant degrees of freedom and form a 
quark-gluon plasma \cite{Hwa:1990xg,Hwa:1995wt}.
\end{itemize}
QCD lattice calculations indicate that the temperature of the transition to the 
quark-gluon plasma is rather low (below 200 MeV \cite{Karsch:2001cy})
as compared to the typical hadronic scale of 1 GeV. Hence one might hope that
the temperature regime II can still be described by a temperature-modified version
of $\chi$PT. In the present work it is shown that this is NOT the case. In addition,
we shall try to determine the temperature which splits regions I and II.

Formally $\chi$PT emerges from a two-step procedure. First, all
other degrees of freedom besides the Goldstone bosons are integrated out. Suppose
for the moment that this can be done --- for the vacuum as well as for the finite
temperature case. The result is a {\it non-local} action, 
schematically\footnote{In the following, we will neglect the generalized mass terms
$\sim \chi, \chi^\dagger$ for simplicity. Their inclusion is straightforward.}
\begin{equation}
  \label{eq:nonlocact}
S = {\rm tr} \int \!\! d^4\! x d^4\!y \, 
D^\mu U(x) \, \Sigma_{\mu\nu}(x-y) \, D^\nu U^\dagger (y) + \ldots
\end{equation}
In a second step the non-local kernels like $\Sigma_{\mu\nu}(x-y)$ have to be expanded
to obtain a {\it local} action/Lagrangian. Technically this can be achieved 
e.g.~by an expansion of the Fourier transform
\begin{equation}
  \label{eq:fouriersigma}
\Sigma_{\mu\nu}(k) = \int \!\! d^4\!(x-y) \, \Sigma_{\mu\nu}(x-y) \, e^{ik(x-y)}
\end{equation}
in powers of momenta $k$. This induces the first term within the trace 
in (\ref{eq:chiptweinb}) and also higher order terms, since in coordinate space each
$k$ translates to a derivative.

Let us first discuss why this two-step procedure works in vacuum. 
The conceptually crucial step is actually the second one.\footnote{In practice, the
first step cannot be carried out due to our limited techniques to solve low-energy QCD.}
To expand (\ref{eq:fouriersigma}) in powers of $k$ requires $\Sigma_{\mu\nu}(k)$ to be
analytic. On the other hand, particle production thresholds cause cuts, 
i.e.~non-analyticities. Fortunately all modes which are integrated out are rather heavy.
Therefore the thresholds for their production are far away from $k=0$. Hence a rather 
broad energy range emerges where vacuum $\chi$PT can yield reliable results.

At finite temperatures the higher excited states (non-Goldstone modes)
are already present in the heat bath. However, at low enough temperatures the
corresponding Boltzmann factors suppress their influence to practically zero.
Temperature regime I emerges. 

Inspired by the success of vacuum $\chi$PT one might
hope that temperature regime II can be similarly described by a modified version
of $\chi$PT:
\begin{equation}
  \label{eq:hope}
{\cal L}(T) \stackrel{?}{=} {1 \over 4} \, {\rm tr}\left(
F^2(T) \, D_\mu U  D^\mu U^\dagger + \tilde F^2(T) \, D_0 U  D_0 U^\dagger
\right) \quad \mbox{+ mass and higher order terms}
\end{equation}
with modified coupling constants, e.g.
\begin{equation}
  \label{eq:ftemp}
F^2(T) = F^2 + O\left( e^{-M_h/T} \right)
\end{equation}
where $M_h$ denotes the mass of the lowest excited state besides the Goldstone modes
and $F^2$ is the vacuum coupling constant from (\ref{eq:chiptweinb}).
The second term within the trace in (\ref{eq:hope}) emerges 
due to the presence of a preferred rest frame (the heat bath):
\begin{equation}
  \label{eq:ftildetemp}
\tilde F^2(T) = O\left( e^{-M_h/T} \right)  \,.
\end{equation}
In practice, it might be possible to obtain at least the lowest order coupling 
constants $F^2(T)$ and $\tilde F^2(T)$ from lattice QCD as 
suggested in \cite{Son:2002ci}.

However, there are two problems with the considerations which have 
led to (\ref{eq:hope}):
First, in general, integrating out degrees of freedom causes 
imaginary parts for the kernels like $\Sigma_{\mu\nu}$ 
(see e.g.~\cite{Greiner:1997dx,Greiner:1998vd} and references therein). 
One can deal with that
complication by a formal doubling of the degrees of freedom, e.g.~by using the real time
formalism \cite{Schwinger:1961qe,Bakshi:1963dv,Bakshi:1963bn,Keldysh:1964ud,%
Chou:1985es,Landsman:1987uw,Das:1997gg}.
The second problem, however, is severe and the present work is
devoted to that problem: If one integrates out specific heavy degrees of freedom at
finite temperature, there appear kernels which are non-analytic at $k=0$. Such terms
obviously invalidate any expansion in terms of local quantities. We should add here
that non-analyticities are connected to thresholds and the latter are connected
to imaginary parts of the kernels. Therefore the two mentioned problems are 
interrelated. Nonetheless we shall concentrate in the present work on the second
problem, i.e.~on non-analytic terms in the real parts of the kernels.

The fact that effective actions at finite temperatures might be manifestly non-local
has been discussed e.g.~in \cite{Weldon:1993bv,Arnold:1993qy,Metikas:1999rm}.
The present work is an application of these findings to 
the case of $\chi$PT.\footnote{We note in passing that thermal non-analyticities within
$\chi$PT have also been discussed e.g.~in \cite{Manuel:1998zk}. That work, however, was
concerned with the derivation of an effective theory for ``soft'' pions by integrating 
out the ``hard'' (=thermal) pions from a non-linear sigma model. In contrast, the 
present work deals with non-Goldstone modes to be integrated out. Here,
``hard'' is the scale of the non-Goldstone modes, while ``soft'' is the scale of
the temperature and the pion mass.}
For simplicity, we restrict ourselves to flavor-$SU(2)$. The pions are the lightest 
Goldstone bosons and therefore the most abandoned ones at finite temperature.
The underlying physical mechanism for the non-analytic terms under consideration is
the Landau damping process: (Virtual) Goldstone bosons scatter with a heavy
mode from the heat bath and form another heavy state. If both heavy states have the
same mass this damping process has its threshold at zero energy and momentum
of the Goldstone boson. In the present work we will identify such processes and
calculate the real parts of the pion self energies which correspond to the Landau
damping mechanism.


A necessary condition for a local expansion of (\ref{eq:nonlocact}) is that the
limit $k \to 0$ is well-defined for (\ref{eq:fouriersigma}). Especially
\begin{equation}
  \label{eq:changelim}
\lim\limits_{\vphantom{\vec k}k_0 \to 0} \, \lim\limits_{\vert \vec k \vert \to 0} 
\Sigma_{\mu\nu}(k) =
\lim\limits_{\vert \vec k \vert \to 0} \, \lim\limits_{\vphantom{\vec k}k_0 \to 0} 
\Sigma_{\mu\nu}(k)  \,.
\end{equation}
We will show in the present work by explicit examples that 
condition (\ref{eq:changelim}) is 
not satisfied for the strong interaction at finite temperature.
Hence there is no local $\chi$PT for the finite temperature regime II.
The big advantage of vacuum $\chi$PT as compared to standard phenomenological
hadronic models is the existence of a systematic power counting scheme. Clearly a
local theory is mandatory for such a scheme. The absence of local $\chi$PT at
finite temperature means that a systematic power counting cannot be developed
--- at least not in the usual way. In other words, we are led back to conventional
hadronic modeling for intermediate temperature low-energy QCD.

The rest of the paper is organized in the following way:
In the next section we will present a non-analytic term appearing in the pion self energy
by coupling the pion to $\rho$- and $\omega$-mesons. In Sec.~\ref{sec:quarks} pions
are coupled to constituent quarks. In Sec.~\ref{sec:sum} we further discuss our results.

\section{Non-analytic term due to interaction with $\rho$- and $\omega$-mesons}
\label{sec:rhoom}

It is the purpose of the present section to pin down the first sign of non-analyticity
when increasing the temperature. Clearly, it has to be connected with the lowest
massive excitations besides the Goldstone bosons. These are the vector mesons
$\rho$ and $\omega$. In the following, we will neglect the very small mass difference 
between $\rho$- and $\omega$-meson and use $M_V \approx 770\,$MeV. 
We will come back to that point in the last section.
Pions can be subject to Landau damping by scattering with one vector meson into
the other one. An appropriate hadronic Lagrangian to describe such a process is
given by \cite{klingl1}
\begin{equation}
  \label{eq:lagrrhopiom}
{\cal L}_{\rm int} = {g_{VVP} \over 4 f_\pi} \, \epsilon^{\mu\nu\alpha\beta} \,
{\rm tr}\left(\partial_\mu V_\nu \, V_\alpha \, \partial_\beta \Phi \right)
\end{equation}
with $g_{VVP} \approx 1.2$ and the pion decay constant $f_\pi \approx 92\,$MeV. $V$
describes the vector mesons and $\Phi$ the Goldstone bosons.

\begin{figure}[htbp]
  \centering
    \includegraphics[keepaspectratio,width=0.7\textwidth]{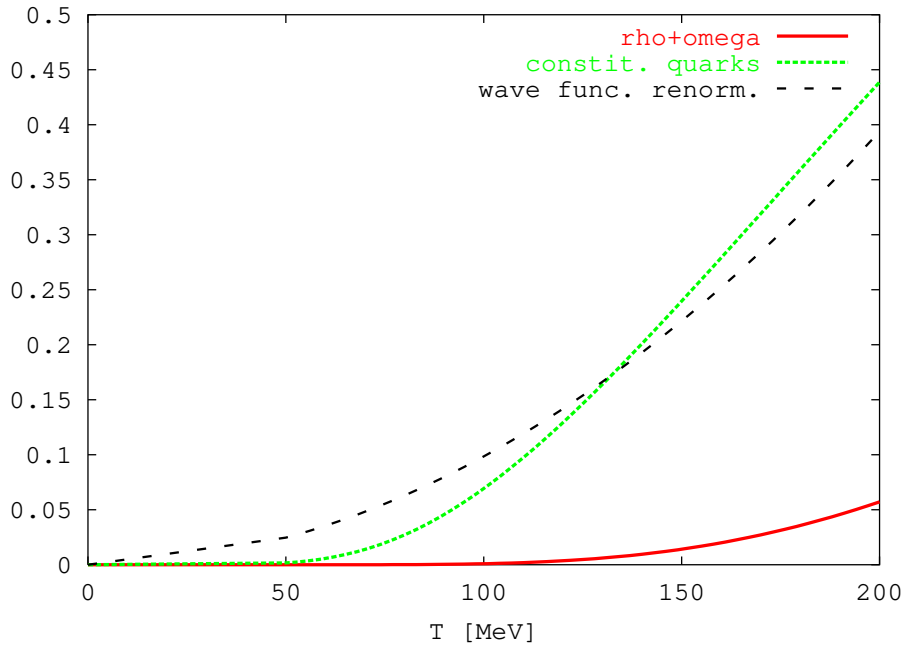}
  \caption{Non-analytic terms emerging from the interaction of pions with $\rho$- and
$\omega$-mesons (full line) and constituent quarks (dotted line). For comparison
the wave function renormalization $Z-1$ caused by thermal pions is also given 
(dashed line). See main text for details.}
  \label{fig:therm}
\end{figure}

In the following we will calculate the thermal one-loop self energy of the $\pi^0$. 
In the loop we have the neutral $\rho$-meson $\rho^0_\mu$
and the $\omega$-meson $\omega_\mu$. The relevant parts of the quantities appearing in
(\ref{eq:lagrrhopiom}) are
\begin{equation}
  \label{eq:phiV}
\Phi \to \left(
  \begin{array}{cc}
    \pi^0 & 0 \\
    0 & -\pi^0
  \end{array}
\right)
\quad , \qquad V_\mu \to \left(
  \begin{array}{cc}
    \rho_\mu^0+\omega_\mu & 0 \\
    0 & -\rho_\mu^0+\omega_\mu
  \end{array}
\right)
\end{equation}
and therefore
\begin{equation}
  \label{eq:lagrneutr}
{\cal L}_{\rm int} \to {g_{VVP} \over f_\pi} \, \epsilon^{\mu\nu\alpha\beta} \,
\rho^0_\mu \, \partial_\nu \omega_\alpha \, \partial_\beta \pi^0  \,.
\end{equation}
The pion self energy is given by
\begin{eqnarray}
  \label{eq:selfenrhom}
\Pi_h(k) = {g^2_{VVP} \over f_\pi^2} \, 
\epsilon^{\mu\nu\alpha\beta} \, \epsilon^{\mu'\nu'\alpha'\beta'} \,
\int \! {d^4 \!p \over (2\pi)^4} \, 
D^\rho_{\mu \mu'}(p-k/2) \, D^\omega_{\alpha \alpha'}(p+k/2)
\, (p+k/2)_\nu \, (p+k/2)_{\nu'} \, k_\beta \, k_{\beta'}
\end{eqnarray}
with
\begin{equation}
  \label{eq:spin1prop}
D^{\rho/\omega}_{\mu \nu}(p) = \left( g_{\mu\nu}-{p_\mu p_\nu \over p^2} \right) \,
D_0(p) + {p_\mu p_\nu \over p^2} \, {1 \over M_V^2}
\end{equation}
and the thermal free scalar boson propagator $D_0$ \cite{Das:1997gg}.
After some algebra one gets
\begin{eqnarray}
  \label{eq:pi1finres}
\Pi_h(k) =  {g^2_{VVP} \over f_\pi^2} \, 2 \, k^2 \int \! {d^4 \!p \over (2\pi)^4} \, 
\left[
\left( M_V^2 + {k^2 \over 4} \right) \, D_0(p-k/2) \, D_0(p+k/2)
- {1 \over 2} \, D_0(p)
\right]  \,.
\end{eqnarray}
As pointed out in the introduction we are interested in the behavior at small $k$.
Here the leading non-analytic term is given by
\begin{equation}
  \label{eq:nonanahad}
{\Pi_h^{\rm non-analyt}(k) \over k^2} = {g^2_{VVP} \over f_\pi^2} \, 2 \, M_V^2 \,
I_B(k)
\end{equation}
with
\begin{equation}
  \label{eq:defib}
I_B(k) = \int \! {d^4 \!p \over (2\pi)^4} \, D_0(p-k/2) \, D_0(p+k/2) - 
\left[\int \! {d^4 \!p \over (2\pi)^4} \, D_0(p-k/2) \, D_0(p+k/2)\right]_{\rm vac} \,.
\end{equation}
The small-$k$ behavior of this integral expression can be immediately taken over from
\cite{Weldon:1993bv}:
\begin{equation}
  \label{eq:reib}
\left.
  \begin{array}{c}
\lim\limits_{\vphantom{\vec k}k_0 \to 0} \lim\limits_{\vert \vec k \vert \to 0} 
{\rm Re}I_B(k) \\[2em]
\lim\limits_{\vert \vec k \vert \to 0} \lim\limits_{\vphantom{\vec k}k_0 \to 0} 
{\rm Re}I_B(k)
  \end{array}
\right\} = {1 \over 4 \pi^2} \int\limits_0^\infty \! {dp \over \omega} \, n_B(\omega)
\left\{
  \begin{array}{c}
    {\displaystyle p^2 \over \displaystyle \omega^2} \\[1.5em] 1
  \end{array}
\right.
\end{equation}
with $\omega=\sqrt{p^2+M_V^2}$ and the Bose distribution
\begin{equation}
  \label{eq:defbose}
n_B(\omega) = {1 \over e^{\,\omega/T} -1}  \,.
\end{equation}
We take the difference between the two different limits given in (\ref{eq:reib})
as a measure for the non-analyticity. We define
\begin{equation}
  \label{eq:defdeltib}
\Delta {\rm Re}I_B = 
{1 \over 4 \pi^2} \int\limits_0^\infty \! {dp \over \omega} \, n_B(\omega) 
\left( 1- {p^2 \over \omega^2} \right) =
{M_V^2 \over 4 \pi^2} \int\limits_0^\infty \! {dp \over \omega^3} \, n_B(\omega) 
\end{equation}
and
\begin{equation}
  \label{eq:defdeltb}
\lim\limits_{k \to 0} 
{\Delta \Pi_h^{\rm non-analyt}(k) \over k^2} = {g^2_{VVP} \over f_\pi^2} \, 2 \, M_V^2 \,
\Delta {\rm Re}I_B  \,.
\end{equation}
The expression (\ref{eq:defdeltb}) is shown in Fig.~\ref{fig:therm} as the full line.
To get an idea about its size relative to other effects appearing at finite temperature
we decided to show in Fig.~\ref{fig:therm} also the effect of the thermal
wave function renormalization for the pion induced by thermal (massless) 
pion loops \cite{Pisarski:1996mt}
\begin{equation}
  \label{eq:wavefr}
  Z-1 = {T^2 \over 12 f_\pi^2} + o(T^4)  \,.
\end{equation}
The expression (\ref{eq:wavefr}) is depicted in 
Fig.~\ref{fig:therm} as the dashed line. We observe that even for temperatures
as high as 200 MeV the non-analytic term due to the $\rho$-$\omega$ loop remains
rather small.
On the other hand, this is probably not the full story about non-analyticity. The vector
mesons are the lightest non-Goldstone degrees of freedom. Therefore they show the
first sign of non-analytic behavior. With rising temperature, however, also more
massive modes become active. The pions might experience Landau damping by
scattering on nucleons and other hadron species. (Note, however, that not all scattering
events have their thresholds at vanishing $k$.) We try to effectively account for
these processes in the next section.

\section{Non-analytic term due to constituent quarks}
\label{sec:quarks}

In the last section we have calculated a Landau damping contribution from scattering
on vector mesons. The utilized Lagrangian (\ref{eq:lagrrhopiom}) is phenomenologically
rather well established. The content of the present section will be a little bit 
more speculative.
Near the chiral transition more and more massive hadronic states become active and
start to interact. One might reinterpret the rising scattering rates as a rising 
movability of the quarks contained in the hadrons. Therefore near the chiral
transition it might be reasonable to describe the system effectively in terms of quarks 
instead of a large bunch of hadrons. Clearly below the transition these quarks have to
be constituent quarks. We use the simplest possible model to describe the interaction
of Goldstone bosons with quarks, the chiral constituent quark model with its 
Lagrangian \cite{diakpet,espraf,Christov:1996vm}
\begin{equation}
  \label{eq:lagrquark}
{\cal L} = \bar q \, \left(i\!\not\!\partial - M U P_L - M U^\dagger P_R \right) \, q
\end{equation}
with the quark fields $q$, the Goldstone bosons encoded in
\begin{equation}
  \label{eq:defU}
U = e^{-i\Phi/f_\pi} \,,
\end{equation}
the projectors on right- and left-handed quarks
\begin{equation}
  \label{eq:defplr}
P_{R,L} = {1 \over 2} (1 \pm \gamma_5)
\end{equation}
and the constituent quark mass $M$.

The self energy for the pion is given by
\begin{eqnarray}
\Pi_q(k) & = & 2 N_c \, {M \over f_\pi^2} \int \! {d^4 \!p \over (2\pi)^4} \, {\rm tr}
(\!\not \!  p  + M) \tilde D_0(p) + 
2 N_c \,  {M^2 \over f_\pi^2} \int \! {d^4 \!p \over (2\pi)^4} \, {\rm tr}
[\gamma_5 \, (\!\not \!  p +M) \, \gamma_5 \, (\!\not \!  k + \!\not \!  p +M)] 
\tilde D_0(p) \, \tilde D_0(k+p)
\nonumber \\
  \label{eq:piq}
& = & 4 N_c \, k^2 \, {M^2 \over f_\pi^2} \int \! {d^4 \!p \over (2\pi)^4} \, 
\tilde D_0(p) \, \tilde D_0(k+p)
\end{eqnarray}
with the number of colors $N_c$ and the free fermion propagator \cite{Das:1997gg}
\begin{equation}
  \label{eq:deffermprop}
(\!\not \!  p +M) \tilde D_0(p)  \,.
\end{equation}
We define in the way completely analogous to Sec.~\ref{sec:rhoom}:
\begin{equation}
  \label{eq:nonanaqu}
{\Pi_q^{\rm non-analyt}(k) \over k^2} = 4 N_c \, {M^2 \over f_\pi^2} \,
I_F(k)
\end{equation}
with
\begin{equation}
  \label{eq:defif}
I_F(k) = \int \! {d^4 \!p \over (2\pi)^4} \, \tilde D_0(p) \, \tilde D_0(p+k) - 
\left[\int \! {d^4 \!p \over (2\pi)^4} \, 
\tilde D_0(p) \, \tilde D_0(p+k)\right]_{\rm vac} \,,
\end{equation}
\begin{equation}
  \label{eq:defdeltif}
\Delta {\rm Re}I_F = 
{M^2 \over 4 \pi^2} \int\limits_0^\infty \! {dp \over \omega^3} \, n_F(\omega) 
\end{equation}
with the Fermi distribution
\begin{equation}
  \label{eq:deffermi}
n_F(\omega) = {1 \over e^{\,\omega/T} +1}
\end{equation}
and finally
\begin{equation}
  \label{eq:defdeltf}
\lim\limits_{k \to 0} 
{\Delta \Pi_q^{\rm non-analyt}(k) \over k^2} = 4 N_c \, {M^2 \over f_\pi^2} \,
\Delta {\rm Re}I_F  \,.
\end{equation}
This last expression is shown in Fig.~\ref{fig:therm} as the dotted line.
For the numerics we have used a vacuum estimate for 
the constituent quark mass: $M \approx 350 \,$MeV.
In principle, one expects that $M$ drops near the chiral transition.
In such a case the result would be even larger as the thermal suppression factor
$e^{-M/T}$ (implicit in (\ref{eq:defdeltif}))
would be less effective. Therefore the dotted line in Fig.~\ref{fig:therm}
provides a conservative estimate of the effect --- provided one accepts on first
place that the chiral constituent quark model yields a reasonable effective 
description of the system. Clearly one should not trust the results down to too low
temperatures: There, thermal excitations have to scale with the Boltzmann factor of
the lowest excitable mode, i.e.~with $e^{-M_V/T}$ and not with $e^{-M/T}$ where
again $M_V$ is the vector-meson mass and $M$ the constituent quark mass. At low
temperatures the lack of confinement of (\ref{eq:lagrquark}) causes non-physical
artifacts. Nonetheless, near the chiral transition the results might be more reasonable.
With all these words of caution in mind we observe that there is a sizable 
non-analyticity induced by the Landau damping on constituent quarks for temperatures
larger than $\approx 100\,$MeV.

\section{Summary}
\label{sec:sum}

The qualitative findings of the presented work and its consequences were
already discussed in Sec.~\ref{sec:intro} and we do not repeat them here.
Quantitatively, we have found that at least for the $\rho$-$\omega$ system
the non-analyticities are rather small. Nonetheless, due to other heavy modes
the effect might be important for temperatures between around 100 MeV and the
chiral transition which is supposed to be somewhat below 200 MeV. We also want to
stress that the discussed non-analyticities cannot be recovered in lattice QCD.
Typical Minkowski-space features like Landau damping are absent in Euclidean-space
calculations. This does not mean that the Euclidean calculations are wrong. It just
tells that analytic continuation can be non-trivial \cite{Weldon:1993bv}.

Coming back
to the $\rho$-$\omega$ system there were two approximations which entered
our calculation: First, strictly speaking the masses of $\rho$ and $\omega$ are
not exactly the same. In principle, this allows for a local expansion around $k=0$.
However, the range of applicability is limited to 
$k^2 < M^2_\omega -M^2_\rho$ \cite{Arnold:1993qy}. 
There is no practical use for such an effective theory. Second, we have neglected
the width of the vector mesons, especially the sizable one of the $\rho$-meson.
It remains to be seen how the inclusion of the widths would influence the results.
This is beyond the scope of the present work.

As already pointed out, the fact that there is no local $\chi$PT at finite temperature 
makes a systematic calculation of a given low-energy quantity much more complicated, 
if not impossible. Especially the interactions of heavier states with themselves
are hard to incorporate 
systematically (in the vacuum this is all integrated out). What can be done for some 
quantities is the following: One might use the Lagrangian of vacuum $\chi$PT with
thermal pions (cf.~the description of temperature regime I above) and in addition a 
non-interacting gas of higher excited states. For the quark condensate this 
approach is utilized in \cite{Gerber:1989tt}. In practice, this might
be sufficient up to temperatures rather close to the chiral 
transition.\footnote{See e.g.~\cite{Karsch:2003zq} for an interesting interpretation 
of lattice data on the thermodynamic quantities of QCD.}
However, first, this approach is not fully systematic (i.e.~it is not clear how to 
calculate corrections) and, second, it is not clear how to perform that for an
arbitrary quantity (say, e.g.~for the pion decay constant).

The Landau damping process, i.e.~here the disappearance of a Goldstone boson 
by scattering
on a heavy state into another heavy mode, can also be reinterpreted as a collective
excitation. In that respect, it is the finite-temperature counterpart of a particle-hole
excitation known from studies of cold nuclear matter \cite{fetterwal}. In other words,
we have discussed the formation of a collective soft mode.
In \cite{Son:2002ci} a local(!) in-medium pion Lagrangian was derived starting from
hydrodynamic considerations. Collective soft modes as caused by Landau damping
processes have not been considered there. It would be interesting to figure out
how the considerations of \cite{Son:2002ci} would be influenced if such collective
states were considered.

\acknowledgments The author thanks M.~Post for a critical reading of the manuscript.
He also thanks U.~Mosel for continuous support.

\bibliography{literature}

\begin{thebibliography}{42}
\expandafter\ifx\csname natexlab\endcsname\relax\def\natexlab#1{#1}\fi
\expandafter\ifx\csname bibnamefont\endcsname\relax
  \def\bibnamefont#1{#1}\fi
\expandafter\ifx\csname bibfnamefont\endcsname\relax
  \def\bibfnamefont#1{#1}\fi
\expandafter\ifx\csname citenamefont\endcsname\relax
  \def\citenamefont#1{#1}\fi
\expandafter\ifx\csname url\endcsname\relax
  \def\url#1{\texttt{#1}}\fi
\expandafter\ifx\csname urlprefix\endcsname\relax\def\urlprefix{URL }\fi
\providecommand{\bibinfo}[2]{#2}
\providecommand{\eprint}[2][]{\url{#2}}

\bibitem[{\citenamefont{Gasser and Leutwyler}(1984)}]{gasleut1}
\bibinfo{author}{\bibfnamefont{J.}~\bibnamefont{Gasser}} \bibnamefont{and}
  \bibinfo{author}{\bibfnamefont{H.}~\bibnamefont{Leutwyler}},
  \bibinfo{journal}{Ann. Phys.} \textbf{\bibinfo{volume}{158}},
  \bibinfo{pages}{142} (\bibinfo{year}{1984}).

\bibitem[{\citenamefont{Gasser and Leutwyler}(1985)}]{gasleut2}
\bibinfo{author}{\bibfnamefont{J.}~\bibnamefont{Gasser}} \bibnamefont{and}
  \bibinfo{author}{\bibfnamefont{H.}~\bibnamefont{Leutwyler}},
  \bibinfo{journal}{Nucl.~Phys.} \textbf{\bibinfo{volume}{B250}},
  \bibinfo{pages}{465, 517, 539} (\bibinfo{year}{1985}).

\bibitem[{\citenamefont{Ecker}(1995)}]{Ecker:1995gg}
\bibinfo{author}{\bibfnamefont{G.}~\bibnamefont{Ecker}},
  \bibinfo{journal}{Prog. Part. Nucl. Phys.} \textbf{\bibinfo{volume}{35}},
  \bibinfo{pages}{1} (\bibinfo{year}{1995}), \eprint{hep-ph/9501357}.

\bibitem[{\citenamefont{Pich}(1995)}]{Pich:1995bw}
\bibinfo{author}{\bibfnamefont{A.}~\bibnamefont{Pich}}, \bibinfo{journal}{Rept.
  Prog. Phys.} \textbf{\bibinfo{volume}{58}}, \bibinfo{pages}{563}
  (\bibinfo{year}{1995}), \eprint{hep-ph/9502366}.

\bibitem[{\citenamefont{Scherer}(2003)}]{Scherer:2002tk}
\bibinfo{author}{\bibfnamefont{S.}~\bibnamefont{Scherer}},
  \bibinfo{journal}{Adv. Nucl. Phys.} \textbf{\bibinfo{volume}{27}},
  \bibinfo{pages}{277} (\bibinfo{year}{2003}), \eprint{hep-ph/0210398}.

\bibitem[{\citenamefont{Creutz}(1983)}]{Creutz:1984mg}
\bibinfo{author}{\bibfnamefont{M.}~\bibnamefont{Creutz}},
  \emph{\bibinfo{title}{Quarks, Gluons and Lattices}}, Cambridge Monographs On
  Mathematical Physics (\bibinfo{publisher}{Cambridge University Press},
  \bibinfo{address}{Cambridge, UK}, \bibinfo{year}{1983}).

\bibitem[{\citenamefont{Giusti et~al.}(2004)\citenamefont{Giusti, Hernandez,
  Laine, Weisz, and Wittig}}]{Giusti:2004yp}
\bibinfo{author}{\bibfnamefont{L.}~\bibnamefont{Giusti}},
  \bibinfo{author}{\bibfnamefont{P.}~\bibnamefont{Hernandez}},
  \bibinfo{author}{\bibfnamefont{M.}~\bibnamefont{Laine}},
  \bibinfo{author}{\bibfnamefont{P.}~\bibnamefont{Weisz}}, \bibnamefont{and}
  \bibinfo{author}{\bibfnamefont{H.}~\bibnamefont{Wittig}},
  \bibinfo{journal}{JHEP} \textbf{\bibinfo{volume}{04}}, \bibinfo{pages}{013}
  (\bibinfo{year}{2004}), \eprint{hep-lat/0402002}.

\bibitem[{\citenamefont{Amoros et~al.}(2001)\citenamefont{Amoros, Bijnens, and
  Talavera}}]{Amoros:2001cp}
\bibinfo{author}{\bibfnamefont{G.}~\bibnamefont{Amoros}},
  \bibinfo{author}{\bibfnamefont{J.}~\bibnamefont{Bijnens}}, \bibnamefont{and}
  \bibinfo{author}{\bibfnamefont{P.}~\bibnamefont{Talavera}},
  \bibinfo{journal}{Nucl. Phys.} \textbf{\bibinfo{volume}{B602}},
  \bibinfo{pages}{87} (\bibinfo{year}{2001}), \eprint{hep-ph/0101127}.

\bibitem[{\citenamefont{Ecker et~al.}(1989{\natexlab{a}})\citenamefont{Ecker,
  Gasser, Pich, and de~Rafael}}]{eckgas}
\bibinfo{author}{\bibfnamefont{G.}~\bibnamefont{Ecker}},
  \bibinfo{author}{\bibfnamefont{J.}~\bibnamefont{Gasser}},
  \bibinfo{author}{\bibfnamefont{A.}~\bibnamefont{Pich}}, \bibnamefont{and}
  \bibinfo{author}{\bibfnamefont{E.}~\bibnamefont{de~Rafael}},
  \bibinfo{journal}{Nucl. Phys.} \textbf{\bibinfo{volume}{B321}},
  \bibinfo{pages}{311} (\bibinfo{year}{1989}{\natexlab{a}}).

\bibitem[{\citenamefont{Ecker et~al.}(1989{\natexlab{b}})\citenamefont{Ecker,
  Gasser, Leutwyler, Pich, and de~Rafael}}]{Ecker:1989yg}
\bibinfo{author}{\bibfnamefont{G.}~\bibnamefont{Ecker}},
  \bibinfo{author}{\bibfnamefont{J.}~\bibnamefont{Gasser}},
  \bibinfo{author}{\bibfnamefont{H.}~\bibnamefont{Leutwyler}},
  \bibinfo{author}{\bibfnamefont{A.}~\bibnamefont{Pich}}, \bibnamefont{and}
  \bibinfo{author}{\bibfnamefont{E.}~\bibnamefont{de~Rafael}},
  \bibinfo{journal}{Phys. Lett.} \textbf{\bibinfo{volume}{B223}},
  \bibinfo{pages}{425} (\bibinfo{year}{1989}{\natexlab{b}}).

\bibitem[{\citenamefont{Donoghue et~al.}(1989)\citenamefont{Donoghue, Ramirez,
  and Valencia}}]{Donoghue:1989ed}
\bibinfo{author}{\bibfnamefont{J.~F.} \bibnamefont{Donoghue}},
  \bibinfo{author}{\bibfnamefont{C.}~\bibnamefont{Ramirez}}, \bibnamefont{and}
  \bibinfo{author}{\bibfnamefont{G.}~\bibnamefont{Valencia}},
  \bibinfo{journal}{Phys. Rev.} \textbf{\bibinfo{volume}{D39}},
  \bibinfo{pages}{1947} (\bibinfo{year}{1989}).

\bibitem[{\citenamefont{Gomez~Nicola and Pelaez}(2002)}]{pelaez02}
\bibinfo{author}{\bibfnamefont{A.}~\bibnamefont{Gomez~Nicola}}
  \bibnamefont{and} \bibinfo{author}{\bibfnamefont{J.~R.}
  \bibnamefont{Pelaez}}, \bibinfo{journal}{Phys. Rev.}
  \textbf{\bibinfo{volume}{D65}}, \bibinfo{pages}{054009}
  (\bibinfo{year}{2002}), \eprint{hep-ph/0109056}.

\bibitem[{\citenamefont{Diakonov and Petrov}(1986)}]{diakpet}
\bibinfo{author}{\bibfnamefont{D.}~\bibnamefont{Diakonov}} \bibnamefont{and}
  \bibinfo{author}{\bibfnamefont{V.~Y.} \bibnamefont{Petrov}},
  \bibinfo{journal}{Nucl. Phys.} \textbf{\bibinfo{volume}{B272}},
  \bibinfo{pages}{457} (\bibinfo{year}{1986}).

\bibitem[{\citenamefont{Espriu et~al.}(1990)\citenamefont{Espriu, de~Rafael,
  and Taron}}]{espraf}
\bibinfo{author}{\bibfnamefont{D.}~\bibnamefont{Espriu}},
  \bibinfo{author}{\bibfnamefont{E.}~\bibnamefont{de~Rafael}},
  \bibnamefont{and} \bibinfo{author}{\bibfnamefont{J.}~\bibnamefont{Taron}},
  \bibinfo{journal}{Nucl. Phys.} \textbf{\bibinfo{volume}{B345}},
  \bibinfo{pages}{22} (\bibinfo{year}{1990}), \bibinfo{note}{erratum-ibid. {\bf
  B355}, 278 (1991)}.

\bibitem[{\citenamefont{Sch\"uren et~al.}(1992)\citenamefont{Sch\"uren,
  Ruiz~Arriola, and Goeke}}]{Schuren:1992sc}
\bibinfo{author}{\bibfnamefont{C.}~\bibnamefont{Sch\"uren}},
  \bibinfo{author}{\bibfnamefont{E.}~\bibnamefont{Ruiz~Arriola}},
  \bibnamefont{and} \bibinfo{author}{\bibfnamefont{K.}~\bibnamefont{Goeke}},
  \bibinfo{journal}{Nucl. Phys.} \textbf{\bibinfo{volume}{A547}},
  \bibinfo{pages}{612} (\bibinfo{year}{1992}).

\bibitem[{\citenamefont{M\"uller and Klevansky}(1994)}]{Mueller:1994dh}
\bibinfo{author}{\bibfnamefont{J.}~\bibnamefont{M\"uller}} \bibnamefont{and}
  \bibinfo{author}{\bibfnamefont{S.~P.} \bibnamefont{Klevansky}},
  \bibinfo{journal}{Phys. Rev.} \textbf{\bibinfo{volume}{C50}},
  \bibinfo{pages}{410} (\bibinfo{year}{1994}).

\bibitem[{\citenamefont{Goity and Leutwyler}(1989)}]{Goity:1989gs}
\bibinfo{author}{\bibfnamefont{J.~L.} \bibnamefont{Goity}} \bibnamefont{and}
  \bibinfo{author}{\bibfnamefont{H.}~\bibnamefont{Leutwyler}},
  \bibinfo{journal}{Phys. Lett.} \textbf{\bibinfo{volume}{B228}},
  \bibinfo{pages}{517} (\bibinfo{year}{1989}).

\bibitem[{\citenamefont{Gerber and Leutwyler}(1989)}]{Gerber:1989tt}
\bibinfo{author}{\bibfnamefont{P.}~\bibnamefont{Gerber}} \bibnamefont{and}
  \bibinfo{author}{\bibfnamefont{H.}~\bibnamefont{Leutwyler}},
  \bibinfo{journal}{Nucl. Phys.} \textbf{\bibinfo{volume}{B321}},
  \bibinfo{pages}{387} (\bibinfo{year}{1989}).

\bibitem[{\citenamefont{Gerber et~al.}(1990)\citenamefont{Gerber, Leutwyler,
  and Goity}}]{Gerber:1990yb}
\bibinfo{author}{\bibfnamefont{P.}~\bibnamefont{Gerber}},
  \bibinfo{author}{\bibfnamefont{H.}~\bibnamefont{Leutwyler}},
  \bibnamefont{and} \bibinfo{author}{\bibfnamefont{J.~L.} \bibnamefont{Goity}},
  \bibinfo{journal}{Phys. Lett.} \textbf{\bibinfo{volume}{B246}},
  \bibinfo{pages}{513} (\bibinfo{year}{1990}).

\bibitem[{\citenamefont{Schenk}(1993)}]{Schenk:1993ru}
\bibinfo{author}{\bibfnamefont{A.}~\bibnamefont{Schenk}},
  \bibinfo{journal}{Phys. Rev.} \textbf{\bibinfo{volume}{D47}},
  \bibinfo{pages}{5138} (\bibinfo{year}{1993}).

\bibitem[{\citenamefont{Hwa}(1990)}]{Hwa:1990xg}
\bibinfo{editor}{\bibfnamefont{R.~C.} \bibnamefont{Hwa}}, ed.,
  \emph{\bibinfo{title}{Quark-gluon plasma}} (\bibinfo{publisher}{World
  Scientific}, \bibinfo{address}{Singapore}, \bibinfo{year}{1990}).

\bibitem[{\citenamefont{Hwa}(1995)}]{Hwa:1995wt}
\bibinfo{editor}{\bibfnamefont{R.~C.} \bibnamefont{Hwa}}, ed.,
  \emph{\bibinfo{title}{Quark-gluon plasma, Vol. 2}} (\bibinfo{publisher}{World
  Scientific}, \bibinfo{address}{Singapore}, \bibinfo{year}{1995}).

\bibitem[{\citenamefont{Karsch}(2002)}]{Karsch:2001cy}
\bibinfo{author}{\bibfnamefont{F.}~\bibnamefont{Karsch}},
  \bibinfo{journal}{Lect. Notes Phys.} \textbf{\bibinfo{volume}{583}},
  \bibinfo{pages}{209} (\bibinfo{year}{2002}), \eprint{hep-lat/0106019}.

\bibitem[{\citenamefont{Son and Stephanov}(2002)}]{Son:2002ci}
\bibinfo{author}{\bibfnamefont{D.~T.} \bibnamefont{Son}} \bibnamefont{and}
  \bibinfo{author}{\bibfnamefont{M.~A.} \bibnamefont{Stephanov}},
  \bibinfo{journal}{Phys. Rev.} \textbf{\bibinfo{volume}{D66}},
  \bibinfo{pages}{076011} (\bibinfo{year}{2002}), \eprint{hep-ph/0204226}.

\bibitem[{\citenamefont{Greiner and M\"uller}(1997)}]{Greiner:1997dx}
\bibinfo{author}{\bibfnamefont{C.}~\bibnamefont{Greiner}} \bibnamefont{and}
  \bibinfo{author}{\bibfnamefont{B.}~\bibnamefont{M\"uller}},
  \bibinfo{journal}{Phys. Rev.} \textbf{\bibinfo{volume}{D55}},
  \bibinfo{pages}{1026} (\bibinfo{year}{1997}), \eprint{hep-th/9605048}.

\bibitem[{\citenamefont{Greiner and Leupold}(1998)}]{Greiner:1998vd}
\bibinfo{author}{\bibfnamefont{C.}~\bibnamefont{Greiner}} \bibnamefont{and}
  \bibinfo{author}{\bibfnamefont{S.}~\bibnamefont{Leupold}},
  \bibinfo{journal}{Annals Phys.} \textbf{\bibinfo{volume}{270}},
  \bibinfo{pages}{328} (\bibinfo{year}{1998}), \eprint{hep-ph/9802312}.

\bibitem[{\citenamefont{Schwinger}(1961)}]{Schwinger:1961qe}
\bibinfo{author}{\bibfnamefont{J.~S.} \bibnamefont{Schwinger}},
  \bibinfo{journal}{J. Math. Phys.} \textbf{\bibinfo{volume}{2}},
  \bibinfo{pages}{407} (\bibinfo{year}{1961}).

\bibitem[{\citenamefont{Bakshi and
  Mahanthappa}(1963{\natexlab{a}})}]{Bakshi:1963dv}
\bibinfo{author}{\bibfnamefont{P.~M.} \bibnamefont{Bakshi}} \bibnamefont{and}
  \bibinfo{author}{\bibfnamefont{K.~T.} \bibnamefont{Mahanthappa}},
  \bibinfo{journal}{J. Math. Phys.} \textbf{\bibinfo{volume}{4}},
  \bibinfo{pages}{1} (\bibinfo{year}{1963}{\natexlab{a}}).

\bibitem[{\citenamefont{Bakshi and
  Mahanthappa}(1963{\natexlab{b}})}]{Bakshi:1963bn}
\bibinfo{author}{\bibfnamefont{P.~M.} \bibnamefont{Bakshi}} \bibnamefont{and}
  \bibinfo{author}{\bibfnamefont{K.~T.} \bibnamefont{Mahanthappa}},
  \bibinfo{journal}{J. Math. Phys.} \textbf{\bibinfo{volume}{4}},
  \bibinfo{pages}{12} (\bibinfo{year}{1963}{\natexlab{b}}).

\bibitem[{\citenamefont{Keldysh}(1964)}]{Keldysh:1964ud}
\bibinfo{author}{\bibfnamefont{L.~V.} \bibnamefont{Keldysh}},
  \bibinfo{journal}{Zh. Eksp. Teor. Fiz.} \textbf{\bibinfo{volume}{47}},
  \bibinfo{pages}{1515} (\bibinfo{year}{1964}), \bibinfo{note}{{Sov.} Phys.
  JETP 20, 1018 (1965)}.

\bibitem[{\citenamefont{Chou et~al.}(1985)\citenamefont{Chou, Su, Hao, and
  Yu}}]{Chou:1985es}
\bibinfo{author}{\bibfnamefont{K.-c.} \bibnamefont{Chou}},
  \bibinfo{author}{\bibfnamefont{Z.-b.} \bibnamefont{Su}},
  \bibinfo{author}{\bibfnamefont{B.-l.} \bibnamefont{Hao}}, \bibnamefont{and}
  \bibinfo{author}{\bibfnamefont{L.}~\bibnamefont{Yu}}, \bibinfo{journal}{Phys.
  Rept.} \textbf{\bibinfo{volume}{118}}, \bibinfo{pages}{1}
  (\bibinfo{year}{1985}).

\bibitem[{\citenamefont{Landsman and van Weert}(1987)}]{Landsman:1987uw}
\bibinfo{author}{\bibfnamefont{N.~P.} \bibnamefont{Landsman}} \bibnamefont{and}
  \bibinfo{author}{\bibfnamefont{C.~G.} \bibnamefont{van Weert}},
  \bibinfo{journal}{Phys. Rept.} \textbf{\bibinfo{volume}{145}},
  \bibinfo{pages}{141} (\bibinfo{year}{1987}).

\bibitem[{\citenamefont{Das}(1997)}]{Das:1997gg}
\bibinfo{author}{\bibfnamefont{A.~K.} \bibnamefont{Das}},
  \emph{\bibinfo{title}{Finite Temperature Field Theory}}
  (\bibinfo{publisher}{World Scientific}, \bibinfo{address}{Singapore},
  \bibinfo{year}{1997}).

\bibitem[{\citenamefont{Weldon}(1993)}]{Weldon:1993bv}
\bibinfo{author}{\bibfnamefont{H.~A.} \bibnamefont{Weldon}},
  \bibinfo{journal}{Phys. Rev.} \textbf{\bibinfo{volume}{D47}},
  \bibinfo{pages}{594} (\bibinfo{year}{1993}).

\bibitem[{\citenamefont{Arnold et~al.}(1993)\citenamefont{Arnold, Vokos,
  Bedaque, and Das}}]{Arnold:1993qy}
\bibinfo{author}{\bibfnamefont{P.}~\bibnamefont{Arnold}},
  \bibinfo{author}{\bibfnamefont{S.}~\bibnamefont{Vokos}},
  \bibinfo{author}{\bibfnamefont{P.~F.} \bibnamefont{Bedaque}},
  \bibnamefont{and} \bibinfo{author}{\bibfnamefont{A.~K.} \bibnamefont{Das}},
  \bibinfo{journal}{Phys. Rev.} \textbf{\bibinfo{volume}{D47}},
  \bibinfo{pages}{4698} (\bibinfo{year}{1993}), \eprint{hep-ph/9211334}.

\bibitem[{\citenamefont{Metikas}(1999)}]{Metikas:1999rm}
\bibinfo{author}{\bibfnamefont{G.}~\bibnamefont{Metikas}}
  (\bibinfo{year}{1999}), \eprint{hep-th/9910063}.

\bibitem[{\citenamefont{Manuel}(1998)}]{Manuel:1998zk}
\bibinfo{author}{\bibfnamefont{C.}~\bibnamefont{Manuel}},
  \bibinfo{journal}{Phys. Rev.} \textbf{\bibinfo{volume}{D57}},
  \bibinfo{pages}{2871} (\bibinfo{year}{1998}), \eprint{hep-ph/9710208}.

\bibitem[{\citenamefont{Klingl et~al.}(1996)\citenamefont{Klingl, Kaiser, and
  Weise}}]{klingl1}
\bibinfo{author}{\bibfnamefont{F.}~\bibnamefont{Klingl}},
  \bibinfo{author}{\bibfnamefont{N.}~\bibnamefont{Kaiser}}, \bibnamefont{and}
  \bibinfo{author}{\bibfnamefont{W.}~\bibnamefont{Weise}}, \bibinfo{journal}{Z.
  Phys.} \textbf{\bibinfo{volume}{A356}}, \bibinfo{pages}{193}
  (\bibinfo{year}{1996}), \eprint{hep-ph/9607431}.

\bibitem[{\citenamefont{Pisarski and Tytgat}(1996)}]{Pisarski:1996mt}
\bibinfo{author}{\bibfnamefont{R.~D.} \bibnamefont{Pisarski}} \bibnamefont{and}
  \bibinfo{author}{\bibfnamefont{M.}~\bibnamefont{Tytgat}},
  \bibinfo{journal}{Phys. Rev.} \textbf{\bibinfo{volume}{D54}},
  \bibinfo{pages}{2989} (\bibinfo{year}{1996}), \eprint{hep-ph/9604404}.

\bibitem[{\citenamefont{Christov et~al.}(1996)}]{Christov:1996vm}
\bibinfo{author}{\bibfnamefont{C.~V.} \bibnamefont{Christov}}
  \bibnamefont{et~al.}, \bibinfo{journal}{Prog. Part. Nucl. Phys.}
  \textbf{\bibinfo{volume}{37}}, \bibinfo{pages}{91} (\bibinfo{year}{1996}),
  \eprint{hep-ph/9604441}.

\bibitem[{\citenamefont{Karsch et~al.}(2003)\citenamefont{Karsch, Redlich, and
  Tawfik}}]{Karsch:2003zq}
\bibinfo{author}{\bibfnamefont{F.}~\bibnamefont{Karsch}},
  \bibinfo{author}{\bibfnamefont{K.}~\bibnamefont{Redlich}}, \bibnamefont{and}
  \bibinfo{author}{\bibfnamefont{A.}~\bibnamefont{Tawfik}},
  \bibinfo{journal}{Phys. Lett.} \textbf{\bibinfo{volume}{B571}},
  \bibinfo{pages}{67} (\bibinfo{year}{2003}), \eprint{hep-ph/0306208}.

\bibitem[{\citenamefont{Fetter and Walecka}(1971)}]{fetterwal}
\bibinfo{author}{\bibfnamefont{A.~L.} \bibnamefont{Fetter}} \bibnamefont{and}
  \bibinfo{author}{\bibfnamefont{J.~D.} \bibnamefont{Walecka}},
  \emph{\bibinfo{title}{Quantum Theory of Many-Particle Systems}}
  (\bibinfo{publisher}{McGraw-Hill}, \bibinfo{address}{New York},
  \bibinfo{year}{1971}).

\end{thebibliography}
\bibliographystyle{apsrev}

\end{document}